\documentclass[english]{article}%
\usepackage{amssymb}
\usepackage{fontenc}
\usepackage[latin1]{inputenc}
\usepackage{amsmath}
\usepackage{fontenc}
\usepackage{babel}
\usepackage{amsfonts}
\usepackage{graphicx}
\usepackage{geometry}%
 
\begin{document}

\title{Quantum states of electromagnetic field interacting with a classical current
and their applications to radiation problems.}
\author{V. G. Bagrov$^{1,2}$\thanks{bagrov@phys.tsu.ru}, D. M. Gitman$^{3,4}%
$\thanks{gitman@if.usp.br}, A. A. Shishmarev$^{2}$%
\thanks{a.a.shishmarev@mail.ru} and A. J. D. Farias Jr$^{4}$%
\thanks{a.jorgedantas@gmail.com}\\$^{1}$ \emph{Department of Physics, Tomsk State University, Lenin Prospekt 36,
634050, Tomsk, Russia;}\\$^{2}$ \emph{Institute of High Current Electronics, SB RAS, Akademichesky Ave.
4, 634055, Tomsk, Russia;}\\$^{3}$ \emph{P.N. Lebedev Physical Institute, 53 Leninskiy prospect, 119991
Moscow, Russia;}\\$^{4}$ \emph{Institute of Physics, University of São Paulo, }\\\emph{Rua do Matão, 1371, CEP 05508-090, São Paulo, SP, Brazil}}
\maketitle

\begin{abstract}
In the beginning, the synchrotron radiation (SR) was studied by classical
methods using the Liénard-Wiechert potentials of electric currents.
Subsequently, quantum corrections to the obtained classical formulas were
studied, considering the emission of photons arising from electronic
transitions between spectral levels, described in terms of the Dirac equation.
In this paper, we consider an intermediate approach, in which electric
currents generating the radiation are considered classically, whereas the
quantum nature of the radiation is taken into account exactly. Such an
approximate approach may be helpful in some cases, it allows one to study the
one-photon and multi-photon radiation without complicating calculations using
corresponding solutions of the Dirac equation. We construct exact quantum
states of the electromagnetic field interacting with classical currents and
study their properties. By their help, we calculate a probability of photon
emission by classical currents and obtain relatively simple formulas for the
one-photon and multi-photon radiation. Using the specific circular electric
current, we calculate the corresponding SR. We discuss a relation of obtained
results with known before, for example, with the Schott formula, with the
Schwinger calculations, with one-photon radiation of scalar particles due to
transitions between Landau levels, and with some previous results of
calculating the two-photon SR.


\end{abstract}

\section{Introduction\label{S1}}

As a rule, a motion of charged particles in external electromagnetic fields is
accompanied by electromagnetic radiation. The most important examples, at the
same time related to the present work, are synchrotron (SR) and cyclotron (CR)
radiations of charged particles in a magnetic field. The phenomenon of SR was
discovered approximately 70 years ago \cite{Elder}. A large number of works
have been devoted to their theoretical description, both within the framework
of classical and quantum theory. In both cases, various approximate methods
and limiting cases were considered. In classical electrodynamics the
electromagnetic field created by an arbitrary electric four-current is
described by potentials of Liénard-Wiechert (LW) \cite{Landau,Jackson}. It
turns out that SR can be described sufficiently precise in the framework of
the classical theory (using LW potentials). Schott was the first to obtain a
successful formula for the angular distribution of power emitted in SR by a
particle moving in a circular orbit \cite{Schott}. An alternative derivation
of classical formulas describing the properties of SR and their deep analysis,
especially for high-energy relativistic electrons, was given by Schwinger
\cite{Schwi49}. Nevertheless, quantum effects may play an important role in
the SR and CR. In particular, effects of a backreaction related to a photon
radiation, aspects of discrete structure of energy levels of electrons in the
magnetic field, and spin properties of charged particles are ignored by the
classical theory. The essence of quantum corrections to classical results was
first pointed out in Ref. \cite{Schwi54}. In quantum theory, radiation rate of
energy of a charge particle in course of quantum transitions was calculated
using exact solutions of the Schrödinger (nonrelativistic case), Klein-Gordon
(spinless case) or Dirac (relativistic case) equations with a magnetic field
\cite{SokolovTernov}. Using his source theory \cite{Schwi} Schwinger had
presented an original derivation of similar results \cite{Schwi73}. Besides,
the quantum treatment revealed a completely new effect of self-polarization of
electrons and positrons moving in uniform and constant magnetic field
\cite{SokTerSelfPol}. We note that in the latter works only a one-photon
radiation in course of quantum transitions was taken into account. However,
there are evidences that a multi-photon emission can contribute significantly
to the SR, see for example \cite{MultiPhot, DualPhot}. For electromagnetic
fields exceeding the critical Schwinger field $H_{0}=m^{2}c^{3}/e\hbar,$
nonlinear phenomena of quantum electrodynamics begin to play a prominent role.
Moreover, at fields comparable with the critical field one can observe
nonlinear quantum effects caused by ultrarelativistic particles with high
enough momenta. Some examples of such effects (of the orders of $\alpha$,
$\alpha^{2}$ ($\alpha$ is a fine-structure constant) in the interaction with
the radiation field) are the one-photon emission by electrons ($e\rightarrow
e\gamma$, $\alpha$), the pair production by photons ($\gamma\rightarrow
e^{+}e^{-}$,$\ \alpha$), electron scattering accompanied by the pair
production ($e\rightarrow ee^{+}e^{-},$ $\alpha^{2}$), two photon emission
process ($e\rightarrow e2\gamma$, $\alpha^{2}$) etc. If an incident particle
has a momentum $p\sim\left(  H_{0}/H\right)  m$, then the probability of the
processes is notable.

It should be noted a significant complexity of calculating even the one-photon
radiation using solutions of the above mentioned quantum equations. There is
an opportunity to simplify these calculations considering in the same
relatively simple manner the multi-photon radiation taking the quantum nature
of the irradiated field into account exactly, but considering the particle
current classically. This means that we neglect the back reaction of the
radiation to the current that generates this radiation. Such an approximation
may be justified in some cases, for example, for high density electron beams.
From the technical point of view, this means that calculating electromagnetic
radiation induced by classical electric currents, we have to work with exact
quantum states of the electromagnetic field interacting with classical
currents. Such an approach is considered in the present work. For these
purposes, we first construct exact quantum states of the electromagnetic field
interacting with classical currents and study their properties. Then by their
help, we calculate a probability of photons emission by a classical current
from the vacuum initial state (i.e., from the state without initial photons).
Then we obtain relatively simple formulas for one-photon and multi-photon
radiation. Using the specific circular electric current, we calculate the
corresponding SR. We discuss a relation of obtained results with known before,
for example, with the Schott formula, with the Schwinger calculations, with
one-photon radiation of scalar particles due to transitions between Landau
levels, and with some known results of calculating the two-photon SR. Less
important technical details are placed in the Appendix.

\section{Quantum states of the radiation field interacting with a classical
current\label{S2}}

Here we consider the quantized electromagnetic field interacting with a
classical current $J_{\mu}\left(  x\right)  $, see Refs.
\cite{Heitl36,Schweber,48,AkhiBer81,GitTy90}. In the Coulomb gauge this system
is described by a Hamiltonian $\hat{H}$ which consists of two terms, a
Hamiltonian of free transversal photons $\hat{H}_{\mathrm{\gamma}}$ and an
interaction Hamiltonian $\hat{H}_{\mathrm{int}}$:
\begin{align}
&  \hat{H}=\hat{H}_{\mathrm{\gamma}}+\hat{H}_{\mathrm{int}},\ \hat
{H}_{\mathrm{\gamma}}=c\hbar\sum_{\lambda=1,2}\int d\mathbf{k}k_{0}\hat
{c}_{\mathbf{k}\lambda}^{\dag}\hat{c}_{\mathbf{k}\lambda}\ ,\nonumber\\
&  \hat{H}_{\mathrm{int}}=\frac{1}{c}\int\left[  J_{i}\left(  x\right)
\hat{A}^{i}\left(  \mathbf{r}\right)  +\frac{1}{2}J_{0}\left(  x\right)
A^{0}\left(  x\right)  \right]  d\mathbf{r}\ . \label{2.1}%
\end{align}
Here $\hat{A}^{i}\left(  \mathbf{r}\right)  $ are operators (in the
Schrödinger representation) of vector potentials of the transversal
electromagnetic field,%
\begin{align}
&  \hat{A}^{i}\left(  \mathbf{r}\right)  =\sqrt{4\pi c\hbar}\sum
\limits_{\lambda=1}^{2}\int d\mathbf{k}\left[  \hat{c}_{\mathbf{k}\lambda
}f_{\mathbf{k}\lambda}^{i}\left(  \mathbf{r}\right)  +\hat{c}_{\mathbf{k}%
\lambda}^{\dag}f_{\mathbf{k}\lambda}^{i\ast}\left(  \mathbf{r}\right)
\right]  ,\ i=1,2,3\ ,\label{2.2}\\
&  f_{\mathbf{k}\lambda}^{i}\left(  \mathbf{r}\right)  =\frac{\exp\left(
i\mathbf{kr}\right)  }{\sqrt{2k_{0}\left(  2\pi\right)  ^{3}}}\epsilon
_{\mathbf{k}\lambda}^{i},\ k_{0}=\left\vert \mathbf{k}\right\vert ,
\label{2.3}%
\end{align}
where $\epsilon_{\mathbf{k}\lambda}^{i}$ are the polarization vectors of the
photon with wave vector $\mathbf{k}$ and polarization $\lambda=1,2$. These
vectors possess the properties
\begin{equation}
\mathbf{\epsilon}_{\mathbf{k}\lambda}\mathbf{\epsilon}_{\mathbf{k}\sigma
}^{\ast}=\delta_{\lambda\sigma},\ \mathbf{\epsilon}_{\mathbf{k}\lambda
}\mathbf{k}=0,\ \sum_{\lambda=1}^{2}\epsilon_{\mathbf{k}\lambda}^{i}%
\epsilon_{\mathbf{k}\lambda}^{j\ast}=\delta^{ij}-\frac{k^{i}k^{j}}{\left\vert
\mathbf{k}\right\vert ^{2}}\ . \label{2.5}%
\end{equation}
Operators $\hat{c}_{\mathbf{k}\lambda}$ and $\hat{c}_{\mathbf{k}\lambda}%
^{\dag}$are the annihilation and creation operators of photons with a wave
vector $\mathbf{k}$ and polarizations $\lambda$. These operators satisfy the
commutation relations:
\begin{equation}
\left[  \hat{c}_{\mathbf{k}\lambda},\hat{c}_{\mathbf{k}^{\prime}%
\lambda^{\prime}}^{\dag}\right]  =\delta_{\lambda\lambda^{\prime}}%
\delta\left(  \mathbf{k-k}^{\prime}\right)  ,\ \left[  \hat{c}_{\mathbf{k}%
\lambda},\hat{c}_{\mathbf{k}^{\prime}\lambda^{\prime}}\right]  =\left[
\hat{c}_{\mathbf{k}\lambda}^{\dag},\hat{c}_{\mathbf{k}^{\prime}\lambda
^{\prime}}^{\dag}\right]  =0\ . \label{2.6}%
\end{equation}
Using Eqs. (\ref{2.2})-(\ref{2.5}) one can verify that the operator $\hat
{A}^{i}\left(  \mathbf{r}\right)  $ satisfy the condition $\operatorname{div}%
\mathbf{\hat{A}}\left(  \mathbf{r}\right)  =0.$ We note that in the Coulomb
gauge $A^{0}\left(  x\right)  $ is a $c$-valued scalar function which
satisfies the following equations:%
\begin{equation}
A^{0}\left(  x\right)  =\int d\mathbf{r}^{\prime}\frac{J_{0}\left(
\mathbf{r}^{\prime},t\right)  }{\left\vert \mathbf{r}-\mathbf{r}^{\prime
}\right\vert },\ \Delta A^{0}\left(  x\right)  =-4\pi J_{0}\left(  x\right)  .
\label{2.8}%
\end{equation}
Then the term $J_{0}\left(  x\right)  A^{0}\left(  x\right)  /2$ can be
represented as $-2\pi J_{0}\left(  x\right)  \Delta^{-1}J_{0}\left(  x\right)
$, and, in general case, is time dependent.

The evolution of state vectors $\left\vert \Psi\left(  t\right)  \right\rangle
$ of the quantized electromagnetic field is governed by the Schrödinger
equation%
\begin{equation}
i\hbar\partial_{t}\left\vert \Psi\left(  t\right)  \right\rangle =\hat
{H}\left\vert \Psi\left(  t\right)  \right\rangle . \label{2.9}%
\end{equation}
The general solution of Eq.~(\ref{2.9}) can be written in the following
form\footnote{See Refs.~\cite{GitBagKuch,BagGitLev}.},
\begin{align}
&  \left\vert \Psi\left(  t\right)  \right\rangle =U\left(  t\right)
\left\vert \Psi\left(  0\right)  \right\rangle ,\label{2.10}\\
&  U\left(  t\right)  =\exp\left[  -i\hbar^{-1}\hat{H}_{\mathrm{\gamma}%
}t\right]  \exp\left[  -i\hbar^{-1}\hat{B}\left(  t\right)  \right]
,\label{2.11}\\
&  \hat{B}\left(  t\right)  =\frac{1}{c}\int_{0}^{t}dt^{\prime}\int\left\{
J_{i}\left(  x^{\prime}\right)  \left[  \hat{A}^{i}\left(  x^{\prime}\right)
+\frac{1}{2}\tilde{A}^{i}\left(  x^{\prime}\right)  \right]  +\frac{1}{2}%
J_{0}\left(  x^{\prime}\right)  A^{0}\left(  x^{\prime}\right)  \right\}
d\mathbf{r}^{\prime},\ \nonumber\\
&  \tilde{A}^{i}\left(  x\right)  =\frac{1}{\hbar c}\int_{0}^{t}dt^{\prime
}\int D_{0}\left(  x-x^{\prime}\right)  \delta_{\bot}^{ik}J^{k}\left(
x^{\prime}\right)  d\mathbf{r}^{\prime},\nonumber\\
&  \hat{A}^{i}\left(  x\right)  =\sqrt{4\pi c\hbar}\sum\limits_{\lambda=1}%
^{2}\int d\mathbf{k}\left[  \hat{c}_{\mathbf{k}\lambda}f_{\mathbf{k}\lambda
}^{i}\left(  x\right)  +\hat{c}_{\mathbf{k}\lambda}^{\dag}f_{\mathbf{k}%
\lambda}^{i\ast}\left(  x\right)  \right]  ,\ f_{\mathbf{k}\lambda}^{i}\left(
x\right)  =f_{\mathbf{k}\lambda}^{i}\left(  \mathbf{r}\right)  e^{-ik_{0}%
ct},\nonumber
\end{align}
where $U\left(  t\right)  $ is an evolution operator, and $\left\vert
\Psi\left(  0\right)  \right\rangle $ is an initial state of the quantized
electromagnetic field at the time instant $t=0$.

The singular function $D_{0}\left(  x-x^{\prime}\right)  $ can be obtained
from the Pauli-Jordan permutation function at $m=0$, see, for example,
\cite{BogolubovShirkov},%
\begin{equation}
\ \square D_{0}\left(  x-x^{\prime}\right)  =0,\ D_{0}\left(  x-x^{\prime
}\right)  =4\pi c\hbar\frac{i}{\left(  2\pi\right)  ^{3}}\int\frac
{d\mathbf{k}}{2k_{0}}\left[  e^{-ik\left(  x-x^{\prime}\right)  }-e^{ik\left(
x-x^{\prime}\right)  }\right]  . \label{2.12}%
\end{equation}
It defines nonequal-time commutation relations for the operators $\hat{A}%
^{i}\left(  x\right)  $,%
\begin{equation}
\left[  \hat{A}^{i}\left(  x\right)  ,\hat{A}^{j}\left(  x^{\prime}\right)
\right]  =-i\delta_{\bot}^{ij}D_{0}\left(  x-x^{\prime}\right)  ,\ \delta
_{\bot}^{ij}=\delta^{ij}-\Delta^{-1}\partial^{i}\partial^{j}, \label{2.13}%
\end{equation}
and is related to the retarded $D^{\mathrm{ret}}\left(  x-x^{\prime}\right)  $
and advanced $D^{\mathrm{adv}}\left(  x-x^{\prime}\right)  $ Green's functions
of the D'Alembert equations,
\begin{align}
&  \int_{0}^{t}dt^{\prime}D_{0}\left(  x-x^{\prime}\right)  =\int_{0}^{\infty
}dt^{\prime}D^{\mathrm{ret}}\left(  x-x^{\prime}\right)  ,\ D_{0}\left(
x-x^{\prime}\right)  =D^{\mathrm{ret}}\left(  x-x^{\prime}\right)
-D^{\mathrm{adv}}\left(  x-x^{\prime}\right)  ,\nonumber\\
&  D^{\mathrm{ret}}\left(  x-x^{\prime}\right)  =\theta\left(  t-t^{\prime
}\right)  D_{0}\left(  x-x^{\prime}\right)  ,\ D^{\mathrm{adv}}\left(
x-x^{\prime}\right)  =\theta\left(  t^{\prime}-t\right)  D_{0}\left(
x-x^{\prime}\right)  ,\nonumber\\
&  \square D^{\mathrm{ret}}\left(  x-x^{\prime}\right)  =\square
D^{\mathrm{adv}}\left(  x-x^{\prime}\right)  =\delta\left(  x-x^{\prime
}\right)  . \label{2.14}%
\end{align}
Taking into account Eqs.~(\ref{2.14}), one can see that the functions
$\tilde{A}^{i}\left(  x\right)  $ represent retarded potentials created by a
classical current (see,\ e.g., \cite{Landau,Galtsov}).

Let us verify directly that state vector (\ref{2.10}) satisfies equation
(\ref{2.9}). Foremost, as the operator $\hat{H}_{\mathrm{\gamma}}$ is
time-independent, we have%
\begin{equation}
i\hbar\partial_{t}\left[  \exp\left(  -i\hbar^{-1}\hat{H}_{\mathrm{\gamma}%
}t\right)  \right]  =\hat{H}_{\mathrm{\gamma}}\exp\left(  -i\hbar^{-1}\hat
{H}_{\mathrm{\gamma}}t\right)  . \label{2.15}%
\end{equation}
However, the derivative $\partial_{t}\hat{A}^{i}\left(  x\right)  $ does not
commute with the operators $\hat{A}^{i}\left(  x^{\prime}\right)
$,\textrm{\ }so when calculating the derivative $i\hbar\partial_{t}$ of the
second exponent in the RHS of Eq. (\ref{2.11}), one has to use the Feynman's
method of disentangling of operators \cite{Fey3}. Calculating the derivative
$i\hbar\partial_{t}$ in such a way, we find%
\begin{align}
&  i\hbar\partial_{t}\exp\left[  -i\hbar^{-1}\hat{B}\left(  t\right)  \right]
=\hat{K}\left(  t\right)  \exp\left[  -i\hat{B}\left(  t\right)  \right]
,\ \hat{K}\left(  t\right)  =\int_{0}^{1}dse^{-is\hbar^{-1}\hat{B}\left(
t\right)  }\left[  \partial_{t}\hat{B}\left(  t\right)  \right]
e^{is\hbar^{-1}\hat{B}\left(  t\right)  },\nonumber\\
&  \ \partial_{t}\hat{B}\left(  t\right)  =\frac{1}{c}\int\left\{
J_{i}\left(  x\right)  \left[  \hat{A}^{i}\left(  x\right)  +\frac{1}{2}%
\tilde{A}^{i}\left(  x\right)  \right]  +\frac{1}{2}J_{0}\left(  x\right)
A^{0}\left(  x\right)  \right\}  d\mathbf{r}. \label{2.17}%
\end{align}
Using the operator relation%
\[
e^{\hat{A}}\hat{M}e^{-\hat{A}}=\hat{M}+\left[  \hat{A},\hat{M}\right]
+\frac{1}{2!}\left[  \hat{A},\left[  \hat{A},\hat{M}\right]  \right]
+\ \ldots\ ,
\]
we represent the integrand in the RHS of $\hat{K}\left(  t\right)  $ as
follows:%
\begin{align}
&  e^{-is\hbar^{-1}\hat{B}\left(  t\right)  }\left[  \partial_{t}\hat
{B}\left(  t\right)  \right]  e^{is\hbar^{-1}\hat{B}\left(  t\right)
}=\partial_{t}\hat{B}\left(  t\right)  +\left[  -is\hbar^{-1}\hat{B}\left(
t\right)  ,\partial_{t}\hat{B}\left(  t\right)  \right] \nonumber\\
&  +\frac{1}{2!}\left[  -is\hbar^{-1}\hat{B}\left(  t\right)  ,\left[
-is\hbar^{-1}\hat{B}\left(  t\right)  ,\partial_{t}\hat{B}\left(  t\right)
\right]  \right]  +\ \ldots\ . \label{2.19}%
\end{align}
Calculating the first commutator in this series, we obtain:%
\begin{equation}
\left[  \hat{B}\left(  t\right)  ,\partial_{t}\hat{B}\left(  t\right)
\right]  =\frac{1}{c^{2}}\int_{0}^{t}dt^{\prime}\int\int\left\{  J_{i}\left(
x^{\prime}\right)  \left[  \hat{A}^{i}\left(  x^{\prime}\right)  ,\hat{A}%
^{j}\left(  x\right)  \right]  J_{j}\left(  x\right)  \right\}  d\mathbf{r}%
d\mathbf{r}^{\prime}. \label{2.20}%
\end{equation}
The nonequal-time commutation relations for the operators $\hat{A}^{i}\left(
x\right)  $ are given by Eq.~(\ref{2.13}). Then (\ref{2.20}) takes the form%
\begin{equation}
\left[  \hat{B}\left(  t\right)  ,\partial_{t}\hat{B}\left(  t\right)
\right]  =-\frac{i}{c^{2}}\int_{0}^{t}dt^{\prime}\int d\mathbf{r}J_{j}\left(
x\right)  \int d\mathbf{r}^{\prime}J_{i}\left(  x^{\prime}\right)
\delta_{\bot}^{ij}D_{0}\left(  x-x^{\prime}\right)  . \label{2.21a}%
\end{equation}
We suppose, as usual, that currents under consideration vanish at spatial
infinities. In this case,%
\begin{equation}
\int d\mathbf{r}^{\prime}J_{i}\left(  x^{\prime}\right)  \delta_{\bot}%
^{ij}D_{0}\left(  x-x^{\prime}\right)  =\int d\mathbf{r}^{\prime}D_{0}\left(
x-x^{\prime}\right)  \delta_{\bot}^{ij}J_{i}\left(  x^{\prime}\right)  .
\label{2.22a}%
\end{equation}
Then, recalling the definition of $\tilde{A}^{i}\left(  x\right)  $ from the
evolution operator (\ref{2.11}), we obtain
\begin{equation}
\left[  \hat{B}\left(  t\right)  ,\partial_{t}\hat{B}\left(  t\right)
\right]  =-\frac{i}{c}\int J_{i}\left(  x\right)  \tilde{A}^{i}\left(
x\right)  d\mathbf{r}. \label{2.23}%
\end{equation}
Since the right-hand side of Eq. (\ref{2.23}) is not an operator, the only
first commutator in the RHS of Eq. (\ref{2.19}) survives. Substituting Eqs.
(\ref{2.17}), (\ref{2.19}) and (\ref{2.23}) into Eq. (\ref{2.17}) and then
integrating over $s$, we find:%
\begin{equation}
\hat{K}\left(  t\right)  =\frac{1}{c}\int\left[  J_{i}\left(  x\right)
\hat{A}^{i}\left(  x\right)  +\frac{1}{2}J_{0}\left(  x\right)  A^{0}\left(
x\right)  \right]  d\mathbf{r}. \label{2.24}%
\end{equation}
Using the fact that in the Coulomb gauge
\begin{equation}
\exp\left[  -i\hbar^{-1}\hat{H}_{\mathrm{\gamma}}t\right]  \hat{K}\left(
t\right)  =\frac{1}{c}\int\left[  J_{i}\left(  x\right)  \hat{A}^{i}\left(
\mathbf{r}\right)  +\frac{1}{2}J_{0}\left(  x\right)  A^{0}\left(  x\right)
\right]  d\mathbf{r}\exp\left[  -i\hbar^{-1}\hat{H}_{\mathrm{\gamma}}t\right]
, \label{2.25}%
\end{equation}
and taking into account Eq. (\ref{2.15}), we make sure that state vector
(\ref{2.10}) does satisfy equation (\ref{2.9}).

It is useful to represent the evolution operator $U\left(  t\right)  $ as:%
\begin{align}
&  U\left(  t\right)  =\exp\left[  i\phi\left(  t\right)  \right]  \exp\left[
-i\hbar^{-1}\hat{H}_{\mathrm{\gamma}}t\right]  \mathcal{D}\left(  y\right)
,\label{2.26}\\
&  \mathcal{D}\left(  y\right)  =\exp\left\{  \sum_{\lambda=1}^{2}\int
d\mathbf{k}\left[  y_{\mathbf{k}\lambda}\left(  t\right)  \hat{c}%
_{\mathbf{k}\lambda}^{\dag}-y_{\mathbf{k}\lambda}^{\ast}\left(  t\right)
\hat{c}_{\mathbf{k}\lambda}\right]  \right\}  ,\label{2.27}\\
&  \phi\left(  t\right)  =-\frac{1}{2c}\int_{0}^{t}dt^{\prime}\int\left[
J_{i}\left(  x^{\prime}\right)  \tilde{A}^{i}\left(  x^{\prime}\right)
+J_{0}\left(  x^{\prime}\right)  A^{0}\left(  x^{\prime}\right)  \right]
d\mathbf{r}^{\prime},\nonumber\\
&  y_{\mathbf{k}\lambda}\left(  t\right)  =-i\sqrt{\frac{4\pi}{\hbar c}}%
\int_{0}^{t}dt^{\prime}\int J_{i}\left(  x^{\prime}\right)  f_{\mathbf{k}%
\lambda}^{i\ast}\left(  x^{\prime}\right)  d\mathbf{r}^{\prime}. \label{2.27a}%
\end{align}
In what follows we omit the argument $\left(  t\right)  $ in functions
$y_{\mathbf{k}\lambda}\left(  t\right)  $ to make formulas more compact.

We remind some basic relations for the displacement operator $\mathcal{D}%
(\alpha)$ in the Coulomb gauge,%
\begin{align}
&  \mathcal{D}^{\dag}(\alpha)=\mathcal{D}^{-1}(\alpha),\ |\alpha
\rangle=\mathcal{D}(\alpha)|0\rangle,\ \hat{c}_{\mathbf{k}\lambda}%
|\alpha\rangle=\alpha_{\mathbf{k}\lambda}|\alpha\rangle,\nonumber\\
&  \mathcal{D}^{\dag}(\alpha)\hat{c}_{\mathbf{k}\lambda}\mathcal{D}%
(\alpha)=\hat{c}_{\mathbf{k}\lambda}+\alpha_{\mathbf{k}\lambda},\ \mathcal{D}%
^{\dag}(\alpha)\hat{c}_{\mathbf{k}\lambda}^{\dag}\mathcal{D}(\alpha)=\hat
{c}_{\mathbf{k}\lambda}^{\dag}+\alpha_{\mathbf{k}\lambda}^{\ast}. \label{2.28}%
\end{align}
With their help, we obtain:
\begin{equation}
\mathcal{D}(y)\left\vert 0\right\rangle =\exp\left(  -\frac{1}{2}\sum
_{\lambda=1}^{2}\int d\mathbf{k\ }\left\vert y_{\mathbf{k}\lambda}\right\vert
^{2}\right)  \exp\left(  \sum_{\lambda=1}^{2}\int d\mathbf{k}\ y_{\mathbf{k}%
\lambda}c_{\mathbf{k}\lambda}^{\dag}\right)  \left\vert 0\right\rangle .
\label{2.28a}%
\end{equation}

\section{Electromagnetic radiation induced by a classical current\label{S3}}

One can use the constructed state vector (\ref{2.10}) to study electromagnetic
radiation induced by a classical current. For simplicity, we choose the vacuum
$\left\vert 0\right\rangle $ as the initial state $\left\vert \Psi\left(
0\right)  \right\rangle $ at the $t=0$ in Eq. (\ref{2.10}). The time evolution
of this initial state follows from the latter equation:%
\begin{equation}
\left\vert \Psi\left(  t\right)  \right\rangle =\exp\left[  i\phi\left(
t\right)  \right]  \exp\left[  -i\hbar^{-1}\hat{H}_{\mathrm{\gamma}}t\right]
\mathcal{D}\left(  y\right)  \left\vert 0\right\rangle . \label{3.1}%
\end{equation}
Using Eq. (\ref{3.1}), we can calculate a probability of photons emission.

When operating in a continuous Fock space, see. \cite{Schweber}, a state with
$N$ photons is formed by the repeated action of the photon creation operators
on the vacuum \textrm{ }$\left\vert 0\right\rangle $, and has the form:%
\begin{equation}
\left\vert \left\{  N\right\}  \right\rangle =\left(  N!\right)  ^{-1/2}%
\prod_{i=1}^{N}\hat{c}_{\mathbf{k}_{i}\lambda_{i}}^{\dag}\left\vert
0\right\rangle , \label{8.0}%
\end{equation}
where $\hat{c}_{\mathbf{k}_{i}\lambda_{i}}^{\dag}$ are creation operators of
photons with wave vector $\mathbf{k}_{i}$ and polarizations $\lambda_{i}$,
$\left\{  N\right\}  =\left(  \mathbf{k}_{1}\lambda_{1},\mathbf{k}_{2}%
\lambda_{2},\ldots,\mathbf{k}_{N}\lambda_{N}\right)  $.

A probability amplitude $R\left(  \left\{  N\right\}  ,t\right)  $ of the
transition from the vacuum state $\left\vert 0\right\rangle $ to the state
(\ref{8.0}) for the time interval $t$ reads:
\begin{equation}
R\left(  \left\{  N\right\}  ,t\right)  =\exp\left[  i\phi\left(  t\right)
\right]  \left\langle 0\right\vert \left(  N!\right)  ^{-1/2}\left(
\prod_{i=1}^{N}\hat{c}_{\mathbf{k}_{i}\lambda_{i}}\right)  \exp\left[
-i\hat{H}_{\mathrm{\gamma}}t\right]  \mathcal{D}\left(  y\right)  \left\vert
0\right\rangle . \label{8.1}%
\end{equation}
Using properties (\ref{2.28}) and (\ref{2.28a}) of the displacement operator
$\mathcal{D}\left(  y\right)  $, and commutation relations (\ref{2.6}), one
can represent amplitude (\ref{8.1}) as follows:
\begin{align}
&  R\left(  \left\{  N\right\}  ,t\right)  =R\left(  0,t\right)  \left(
N!\right)  ^{-1/2}\prod_{i=1}^{N}\exp\left[  -i\left\vert \mathbf{k}%
_{i}\right\vert ct\right]  y_{\mathbf{k}_{i}\lambda_{i}},\nonumber\\
&  R\left(  0,t\right)  =\langle0\left\vert \Psi\left(  t\right)
\right\rangle =\exp\left[  i\phi\left(  t\right)  \right]  \exp\left(
-\frac{1}{2}\sum_{\lambda=1}^{2}\int d\mathbf{k}\left\vert y_{\mathbf{k}%
\lambda}\right\vert ^{2}\right)  . \label{8.2}%
\end{align}
Then the corresponding differential probability $P\left(  \left\{  N\right\}
,t\right)  $ of such a transition (which we interpret as differential
probability of the photon emission) has the form:%
\begin{align}
&  P\left(  \left\{  N\right\}  ,t\right)  =\left\vert R\left(  \left\{
N\right\}  ,t\right)  \right\vert ^{2}=p\left(  \left\{  N\right\}  ,t\right)
P\left(  0,t\right)  ,\nonumber\\
&  p\left(  \left\{  N\right\}  ,t\right)  =\left(  N!\right)  ^{-1}%
\prod_{i=1}^{N}\left\vert y_{\mathbf{k}_{i}\lambda_{i}}\right\vert
^{2},\nonumber\\
&  P\left(  0,t\right)  =|R\left(  0,t\right)  |^{2}=\exp\left(
-\sum_{\lambda=1}^{2}\int d\mathbf{k}\left\vert y_{\mathbf{k}\lambda
}\right\vert ^{2}\right)  , \label{8.2a}%
\end{align}
where\emph{ }$P\left(  0,t\right)  $\emph{ }is the\emph{ }vacuum-to-vacuum
transition probability, or the probability of a transition without any photon
emission. Thus,$\ p\left(  \left\{  N\right\}  ,t\right)  $ is the relative
probability of a process in which $N$ photons with quantum numbers
$\mathbf{k}_{i}\lambda_{i}$ are emitted (the relative differential probability).

One can obtain the total probability $P\left(  N,t\right)  $ of transition
from the vacuum state $\left\vert 0\right\rangle $ to the state with $N$
arbitrary photons, summing the quantity $p\left(  \left\{  N\right\}
,t\right)  $ over the sets $\left\{  N\right\}  $. Thus, we get\footnote{It
should be noted that Glauber \cite{glauber} derived the total probability
$P\left(  N,t\right)  $ by his own method, however, did not consider its
application for the radiation problem.}:
\begin{align}
&  P\left(  N,t\right)  =\sum_{\left\{  N\right\}  }P\left(  \left\{
N\right\}  ,t\right)  =P\left(  0,t\right)  p\left(  N,t\right)
,\ \sum_{\left\{  N\right\}  }=\prod_{i=1}^{N}\left(  \sum_{\lambda_{i}}\int
d\mathbf{k}_{i}\right)  ,\nonumber\\
&  p\left(  N,t\right)  =\left(  N!\right)  ^{-1}\prod_{i=1}^{N}\left(
\sum_{\lambda_{i}}\int d\mathbf{k}_{i}\left\vert y_{\mathbf{k}_{i}\lambda_{i}%
}\right\vert ^{2}\right)  . \label{8.10}%
\end{align}

Introducing a total probability $P\left(  t\right)  $ of the photon emission
for the time interval $t$ as follows
\begin{equation}
P\left(  t\right)  =\sum_{N=1}^{\infty}P\left(  N,t\right)  =P\left(
0,t\right)  \sum_{N=1}^{\infty}\left(  N!\right)  ^{-1}\prod_{i=1}^{N}\left(
\sum_{\lambda_{i}}\int d\mathbf{k}_{i}\left\vert y_{\mathbf{k}_{i}\lambda_{i}%
}\right\vert ^{2}\right)  , \label{8.3}%
\end{equation}
one can easily verify that the relation $P\left(  0,t\right)  +P\left(
t\right)  =1$ holds true.

The electromagnetic energy of $\left\{  N\right\}  $ photons with given
quantum numbers $\left\{  \mathbf{k}\lambda\right\}  =\left(  \mathbf{k}%
_{i}\lambda_{i},i=1,2,\ldots,N\right)  $ depends only on their momenta
$\left\{  \mathbf{k}\right\}  =\left(  \mathbf{k}_{i},i=1,2,\ldots,N\right)  $
and does not depend on their polarizations; it is equal to%
\begin{equation}
W\left(  \left\{  N\right\}  \right)  =\hbar c\left[  \sum_{i=1}^{N}\left\vert
\mathbf{k}_{i}\right\vert \right]  . \label{8.11}%
\end{equation}
Then the total electromagnetic energy $W\left(  N,t\right)  $ of $N$ emitted
photon reads:%
\begin{equation}
W\left(  N,t\right)  =\sum_{\left\{  N\right\}  }W\left(  \left\{  N\right\}
\right)  p\left(  \left\{  N\right\}  ,t\right)  =\hbar c\left(  N!\right)
^{-1}\sum_{\lambda_{1}=1}^{2}\sum_{\lambda_{2}=1}^{2}\ldots\sum_{\lambda
_{N}=1}^{2}\int d\mathbf{k}_{1}d\mathbf{k}_{2}\ldots d\mathbf{k}_{N}\left[
\sum_{j=1}^{N}\left\vert \mathbf{k}_{i}\right\vert \right]  \prod_{i=1}%
^{N}\left\vert y_{\mathbf{k}_{i}\lambda_{i}}\right\vert ^{2}. \label{8.12}%
\end{equation}
It is easy to demonstrate (see Appendix) that $W\left(  N,t\right)  $ can be
represented as\textrm{:}%
\begin{align}
&  W\left(  N,t\right)  =\frac{A}{\left(  N-1\right)  !}\left(  \sum
_{\lambda=1}^{2}\int d\mathbf{k}\left\vert y_{\mathbf{k}\lambda}\right\vert
^{2}\right)  ^{N-1},\nonumber\\
&  A=\hbar c\sum_{\lambda=1}^{2}\int d\mathbf{k}k_{0}\left\vert y_{\mathbf{k}%
\lambda}\right\vert ^{2},\ k_{0}=\left\vert \mathbf{k}\right\vert .
\label{8.13}%
\end{align}

Finally, we calculate the total energy $W\left(  t\right)  $ of emitted
photons:%
\begin{equation}
W\left(  t\right)  =\sum_{N=1}^{\infty}W\left(  N,t\right)  . \label{8.14}%
\end{equation}
The sum (\ref{8.14}) can be calculated exactly, taking into account Eq.
(\ref{8.13}),
\begin{equation}
W\left(  t\right)  =A\exp\sum_{\lambda=1}^{2}\int d\mathbf{k}\left\vert
y_{\mathbf{k}\lambda}\right\vert ^{2}. \label{8.15}%
\end{equation}

\section{One-photon radiation by a circular current\label{S3.1}}

Here we study one-photon radiation from the vacuum induced by a specific
circular current. Here we are interested in calculating one-photon radiation,
that is why we will discuss a probability of the appearance of one photon with
given quantum numbers $\mathbf{k}$ and $\lambda=1,2$. Thus, we consider a
transition amplitude from the state (\ref{3.1}) to the final state of the form
(\ref{8.0}) with $N=1$.\textrm{ }Using (\ref{8.12}), we write one-photon
emission as%
\begin{equation}
W\left(  1,t\right)  =\hbar c\sum_{\lambda=1}^{2}\int d\mathbf{k}%
k_{0}\left\vert y_{\mathbf{k}\lambda}\right\vert ^{2},\ k_{0}=\left\vert
\mathbf{k}\right\vert . \label{3.12}%
\end{equation}

Let us consider a circular current formed by electrons moving perpendicularly
to an external uniform and constant magnetic field $\mathbf{H}=\left(
0,0,H\right)  $ with the velocity $\mathbf{v}$ along a circular trajectory of
the radius $R$. Such a current has the following form \cite{SokolovTernov}:%
\begin{align}
&  J_{0}\left(  x\right)  =q\delta^{\left(  3\right)  }\left(  \mathbf{r}%
-\mathbf{r}\left(  t\right)  \right)  ,\ \mathbf{J}\left(  x\right)
=q\mathbf{\dot{r}}\left(  t\right)  \delta^{\left(  3\right)  }\left(
\mathbf{r}-\mathbf{r}\left(  t\right)  \right)  ,\nonumber\\
&  \mathbf{r}\left(  t\right)  =\left(  R\cos\omega t,R\sin\omega t,0\right)
,\ \mathbf{v}\left(  t\right)  =\mathbf{\dot{r}}\left(  t\right)  =\omega
R\left(  -\sin\omega t,\cos\omega t,0\right)  , \label{4.1}%
\end{align}
where $q=-e$, $e>0$ is the electron charge, $\omega=eH/mc$ is the cyclotron
frequency. We disregard the backreaction of the radiation, i.e., we suppose
that the current is maintained in its original form during the time interval
$\Delta t=t$.

Functions $y_{\mathbf{k}\lambda}$ (\ref{2.26}) for the current (\ref{4.1})
have the form:%
\begin{align}
&  y_{\mathbf{k}\lambda}=iq%
%TCIMACRO{\dint \limits_{0}^{t}}%
%BeginExpansion
{\displaystyle\int\limits_{0}^{t}}
%EndExpansion
dt^{\prime}\frac{\mathbf{v}\left(  t^{\prime}\right)  \mathbf{\epsilon
}_{\mathbf{k}\lambda}^{\ast}}{\sqrt{\hbar ck_{0}\left(  2\pi\right)  ^{2}}%
}\exp\left\{  i\left[  k_{0}ct^{\prime}-\mathbf{kr}\left(  t^{\prime}\right)
\right]  \right\}  ,\label{4.4}\\
&  \mathbf{k}=\left(  k_{\perp}\cos\varphi,k_{\perp}\sin\varphi,k_{\Vert
}\right)  ,\ k_{\perp}=k_{0}\sin\theta,\ k_{\Vert}=k_{0}\cos\theta.
\label{4.5}%
\end{align}
Here $\varphi$ is the angle between the $x$ axis and the projection of the
vector $\mathbf{k}$ onto the $xy$ plane, and $\theta$ is the angle between the
$z$ axis and $\mathbf{k}$. Thus,%
\begin{equation}
W\left(  1,t\right)  =\frac{\hbar c}{\left(  2\pi\right)  ^{2}}\int
d\mathbf{k}\ k_{0}\left\vert \int dt^{\prime}\mathbf{J}\left(  x^{\prime
}\right)  \mathbf{\epsilon}_{\mathbf{k}\lambda}^{\ast}\exp\left[
ik_{0}ct^{\prime}-\mathbf{kr}\left(  t^{\prime}\right)  \right]  \right\vert
^{2}. \label{9.4}%
\end{equation}
Then
\begin{align}
&  \exp\left[  -i\mathbf{kr}\left(  t^{\prime}\right)  \right]  =\exp\left[
-ik_{\perp}R\sin\tau\right]  ,\nonumber\\
&  \exp\left(  ik_{0}ct^{\prime}\right)  =\exp\left[  ick_{0}\omega
^{-1}\left(  \varphi-\pi/2\right)  \right]  \exp\left(  ick_{0}\omega^{-1}%
\tau\right)  ,\nonumber\\
&  \mathbf{v}\left(  \tau\right)  =\omega R\left[  \cos\left(  \tau
+\varphi\right)  ,\sin\left(  \tau+\varphi\right)  ,0\right]  ,\nonumber\\
&  \tau=\tau_{\mathrm{i}}+\omega t^{\prime},\ \tau_{\mathrm{i}}=\pi
/2-\varphi,\
%TCIMACRO{\dint \limits_{0}^{t}}%
%BeginExpansion
{\displaystyle\int\limits_{0}^{t}}
%EndExpansion
dt^{\prime}\rightarrow%
%TCIMACRO{\dint \limits_{\tau_{\mathrm{i}}}^{\tau_{\mathrm{i}}+\omega t}}%
%BeginExpansion
{\displaystyle\int\limits_{\tau_{\mathrm{i}}}^{\tau_{\mathrm{i}}+\omega t}}
%EndExpansion
\omega^{-1}d\tau\ . \label{4.8}%
\end{align}

In the case under consideration, we chose linear polarization vectors
$\mathbf{\epsilon}_{\mathbf{k}\lambda}$ as:
\begin{align}
&  \mathbf{\epsilon}_{\mathbf{k}1}=\left(  \cos\varphi\cos\theta,\sin
\varphi\cos\theta,-\sin\theta\right)  ,\ \mathbf{\epsilon}_{\mathbf{k}%
2}=\left(  -\sin\varphi,\cos\varphi,0\right)  ,\nonumber\\
&  \mathbf{\epsilon}_{\mathbf{k}1}\mathbf{\epsilon}_{\mathbf{k}1}%
=\mathbf{\epsilon}_{\mathbf{k}2}\mathbf{\epsilon}_{\mathbf{k}2}%
=1,\ \mathbf{\epsilon}_{\mathbf{k}1}\mathbf{\epsilon}_{\mathbf{k}%
2}=\mathbf{\epsilon}_{\mathbf{k}1}\mathbf{k}=\mathbf{\epsilon}_{\mathbf{k}%
2}\mathbf{k}=0. \label{4.11}%
\end{align}
One can easily verify that the following relations hold:%
\begin{equation}
\mathbf{v}\left(  t^{\prime}\right)  \mathbf{\epsilon}_{\mathbf{k}1}^{\ast
}=\omega R\cos\theta\cos\tau,\ \mathbf{v}\left(  t^{\prime}\right)
\mathbf{\epsilon}_{\mathbf{k}2}^{\ast}=\omega R\sin\tau. \label{4.12}%
\end{equation}
Now it follows from Eqs.~(\ref{4.4}) that
\begin{align}
&  y_{\mathbf{k}1}=\frac{iqR\cos\theta}{\sqrt{k_{0}\left(  2\pi\right)
^{2}\hbar c}}Y_{\mathbf{k}}\left(  \varphi\right)
%TCIMACRO{\dint \limits_{\tau_{\mathrm{i}}}^{\tau_{\mathrm{i}}+\omega t}}%
%BeginExpansion
{\displaystyle\int\limits_{\tau_{\mathrm{i}}}^{\tau_{\mathrm{i}}+\omega t}}
%EndExpansion
d\tau\exp\left(  ick_{0}\omega^{-1}\tau\right)  \cos\tau\exp\left(
-ik_{\perp}R\sin\tau\right)  ,\nonumber\\
&  y_{\mathbf{k}2}=\frac{iqR}{\sqrt{k_{0}\left(  2\pi\right)  ^{2}\hbar c}%
}Y_{\mathbf{k}}\left(  \varphi\right)
%TCIMACRO{\dint \limits_{\tau_{\mathrm{i}}}^{\tau_{\mathrm{i}}+\omega t}}%
%BeginExpansion
{\displaystyle\int\limits_{\tau_{\mathrm{i}}}^{\tau_{\mathrm{i}}+\omega t}}
%EndExpansion
d\tau\exp\left(  ick_{0}\omega^{-1}\tau\right)  \sin\tau\exp\left(
-ik_{\perp}R\sin\tau\right)  ,\nonumber\\
&  Y_{\mathbf{k}}\left(  \varphi\right)  =\exp\left[  ick_{0}\omega
^{-1}\left(  \varphi-\pi/2\right)  \right]  . \label{4.13}%
\end{align}

At this stage, we utilize a well-known plane wave expansion of the Bessel
functions $j_{n}\left(  x\right)  $ (see, e.g., \cite{SokolovTernov}),
\begin{align}
&  \exp\left(  -ik_{\perp}R\sin\tau\right)  =\sum_{n=-\infty}^{+\infty}%
j_{n}\left(  k_{\perp}R\right)  \exp\left(  -in\tau\right)  ,\nonumber\\
&  \sin\tau\exp\left(  -ik_{\perp}R\sin\tau\right)  =i\sum_{n=-\infty
}^{+\infty}j_{n}^{\prime}\left(  k_{\perp}R\right)  \exp\left(  -in\tau
\right)  ,\nonumber\\
&  \cos\tau\exp\left(  -ik_{\perp}R\sin\tau\right)  =\sum_{n=-\infty}%
^{+\infty}\frac{n}{k_{\perp}R}j_{n}\left(  k_{\perp}R\right)  \exp\left(
-in\tau\right)  . \label{4.14}%
\end{align}
Using (\ref{4.14}) in Eqs.~(\ref{4.13}), we obtain:%
\begin{align}
&  y_{\mathbf{k}1}=i\frac{qR\cos\theta}{\sqrt{k_{0}\left(  2\pi\right)
^{2}\hbar c}}Y_{\mathbf{k}}\left(  \varphi\right)  \sum_{n=-\infty}^{+\infty
}\frac{nj_{n}\left(  k_{\perp}R\right)  }{k_{\perp}R}F_{\mathbf{k}}^{n}\left(
\varphi,t\right)  ,\nonumber\\
&  y_{\mathbf{k}2}=-\frac{qR}{\sqrt{k_{0}\left(  2\pi\right)  ^{2}\hbar c}%
}Y_{\mathbf{k}}\left(  \varphi\right)  \sum_{n=-\infty}^{+\infty}j_{n}%
^{\prime}\left(  k_{\perp}R\right)  F_{\mathbf{k}}^{n}\left(  \varphi
,t\right)  ,\nonumber\\
&  F_{\mathbf{k}}^{n}\left(  \varphi,t\right)  =%
%TCIMACRO{\dint \limits_{\tau_{\mathrm{i}}}^{\tau_{\mathrm{i}}+\omega t}}%
%BeginExpansion
{\displaystyle\int\limits_{\tau_{\mathrm{i}}}^{\tau_{\mathrm{i}}+\omega t}}
%EndExpansion
d\tau\exp\left[  i\left(  ck_{0}\omega^{-1}-n\right)  \tau\right]  ,
\label{4.15}%
\end{align}
we can rewrite (\ref{4.15}) as follows:\
\begin{align}
&  y_{\mathbf{k}1}=\frac{iq\cot\theta}{\sqrt{k_{0}^{3}\left(  2\pi\right)
^{2}\hbar c}}Y_{\mathbf{k}}\left(  \varphi\right)  \sum_{n=-\infty}^{+\infty
}nj_{n}\left(  k_{\perp}R\right)  F_{\mathbf{k}}^{n}\left(  \varphi,t\right)
,\nonumber\\
&  y_{\mathbf{k}2}=-\frac{qR}{\sqrt{k_{0}\left(  2\pi\right)  ^{2}\hbar c}%
}Y_{\mathbf{k}}\left(  \varphi\right)  \sum_{n=-\infty}^{+\infty}j_{n}%
^{\prime}\left(  k_{\perp}R\right)  F_{\mathbf{k}}^{n}\left(  \varphi
,t\right)  . \label{4.17}%
\end{align}
Now, we can calculate the corresponding probabilities $\left\vert
y_{\mathbf{k}\lambda}\right\vert ^{2}$,
\begin{align}
&  \ \left\vert y_{\mathbf{k}1}\right\vert ^{2}=\frac{q^{2}}{\hbar c}%
\frac{\cot^{2}\theta}{k_{0}^{3}\left(  2\pi\right)  ^{2}}\left\vert
\sum_{n=-\infty}^{+\infty}nj_{n}\left(  k_{\perp}R\right)  F_{\mathbf{k}}%
^{n}\left(  \varphi,t\right)  \right\vert ^{2},\nonumber\\
&  \ \left\vert y_{\mathbf{k}2}\right\vert ^{2}=\frac{q^{2}}{\hbar c}%
\frac{R^{2}}{k_{0}\left(  2\pi\right)  ^{2}}\left\vert \sum_{n=-\infty
}^{+\infty}j_{n}^{\prime}\left(  k_{\perp}R\right)  F_{\mathbf{k}}^{n}\left(
\varphi,t\right)  \right\vert ^{2}. \label{4.18}%
\end{align}

The radiated energy (\ref{3.12}) has to be calculated in the following manner:%
\begin{align}
&  W\left(  1,t\right)  =W_{1}\left(  1,t\right)  +W_{2}\left(  1,t\right)
,\nonumber\\
&  W_{1}\left(  1,t\right)  =\hbar c\int d\mathbf{k}k_{0}\left\vert
y_{\mathbf{k}1}\right\vert ^{2}=\int d\mathbf{k}\frac{q^{2}\cot^{2}\theta
}{k_{0}^{2}\left(  2\pi\right)  ^{2}}\left\vert \sum_{n=-\infty}^{+\infty
}nj_{n}\left(  k_{\perp}R\right)  F_{\mathbf{k}}^{n}\left(  \varphi,t\right)
\right\vert ^{2},\nonumber\\
&  W_{2}\left(  1,t\right)  =\hbar c\int d\mathbf{k}k_{0}\left\vert
y_{\mathbf{k}2}\right\vert ^{2}=\int d\mathbf{k}\frac{q^{2}R^{2}}{\left(
2\pi\right)  ^{2}}\left\vert \sum_{n=-\infty}^{+\infty}j_{n}^{\prime}\left(
k_{\perp}R\right)  F_{\mathbf{k}}^{n}\left(  \varphi,t\right)  \right\vert
^{2}. \label{4.20}%
\end{align}
Note that the functions $F_{\mathbf{k}}^{n}\left(  \varphi,t\right)  $ can be
represented as:
\begin{equation}
F_{\mathbf{k}}^{n}\left(  \varphi,t\right)  =\omega\exp\left[  -i\left(
ck_{0}\omega^{-1}-n\right)  \varphi\right]  \exp\left[  i\frac{\pi}{2}\left(
ck_{0}\omega^{-1}-n\right)  \right]
%TCIMACRO{\dint \limits_{0}^{t}}%
%BeginExpansion
{\displaystyle\int\limits_{0}^{t}}
%EndExpansion
dt^{\prime}\exp\left[  i\left(  ck_{0}-n\omega\right)  t^{\prime}\right]  .
\label{4.21}%
\end{equation}
Using the well-known integral representation of Kronecker's delta function
\begin{equation}
\oint d\varphi\exp\left[  i\left(  n-n^{\prime}\right)  \varphi\right]
=2\pi\delta_{nn\prime}, \label{4.22}%
\end{equation}
we can transform the quantities $W_{1}\left(  1,t\right)  $ and $W_{2}\left(
1,t\right)  $ as follows:%
\begin{align}
&  W_{1}\left(  1,t\right)  =q^{2}\omega^{2}\sum_{n=-\infty}^{+\infty}\int
_{0}^{\infty}\frac{dk_{0}}{2\pi}\int_{0}^{\pi}\sin\theta d\theta\ \cot
^{2}\theta\ n^{2}j_{n}^{2}\left(  k_{\perp}R\right)  \left\vert \int_{0}%
^{t}dt^{\prime}\ \exp\left[  i\left(  ck_{0}-n\omega\right)  t^{\prime
}\right]  \right\vert ^{2},\nonumber\\
&  W_{2}\left(  1,t\right)  =q^{2}\omega^{2}R^{2}\sum_{n=-\infty}^{+\infty
}\int_{0}^{\infty}\frac{dk_{0}}{2\pi}\int_{0}^{\pi}\sin\theta d\theta
\ k_{0}^{2}\ j_{n}^{\prime2}\left(  k_{\perp}R\right)  \left\vert \int_{0}%
^{t}dt^{\prime}\ \exp\left[  i\left(  ck_{0}-n\omega\right)  t^{\prime
}\right]  \right\vert ^{2}. \label{4.23}%
\end{align}
Then the the energy $W\left(  1,t\right)  $ reads:%
\begin{equation}
W\left(  1,t\right)  =\frac{q^{2}\omega^{2}}{2\pi}\sum_{n=-\infty}^{+\infty
}\int_{0}^{\infty}\!dk_{0}\!\int_{0}^{\pi}\sin\theta d\theta\left[
n^{2}\ j_{n}^{2}\left(  k_{\perp}R\right)  \cot^{2}\theta+k_{0}^{2}%
\ R^{2}\ j_{n}^{\prime2}\left(  k_{\perp}R\right)  \right]  \left\vert
\int_{0}^{t}dt^{\prime} \exp\left[  i\left(  ck_{0}-n\omega\right)  t^{\prime
}\right]  \right\vert ^{2}\!\!. \label{4.24}%
\end{equation}

\subsection{Derivation of the Schott formula}

Let us study the time behavior of the energy $W\left(  1,t\right)  $ of the
one-photon emission (\ref{4.24}). One can see that at $t\rightarrow\infty$
this quantity as a function of time is not well defined. However, a real
physical meaning has the rate $w\left(  t\right)  $ of the energy emission,
which is the time derivative of $W\left(  1,t\right)  $,%
\begin{align}
&  w\left(  t\right)  =\partial_{t}W\left(  1,t\right)  =\frac{q^{2}\omega
^{2}}{2\pi}\sum_{n=-\infty}^{+\infty}K\left(  t\right)  \int_{0}^{\infty
}dk_{0}\int_{0}^{\pi}\sin\theta\left[  n^{2}j_{n}^{2}\left(  k_{\perp
}R\right)  \cot^{2}\theta+k_{0}^{2}R^{2}j_{n}^{\prime2}\left(  k_{\perp
}R\right)  \right]  d\theta,\nonumber\\
&  K\left(  t\right)  =\frac{\partial}{\partial t}\left\vert \int_{0}%
^{t}dt^{\prime}\ \exp\left[  i\left(  ck_{0}-n\omega\right)  t^{\prime
}\right]  \right\vert ^{2}. \label{4.32}%
\end{align}
To compare with the Schott result, we have to consider $w\left(  t\right)  $
as $t\rightarrow\infty.$ In fact the problem is reduced to calculating the
$\lim_{t\rightarrow\infty}K\left(  t\right)  .$ This limit can be easily
calculated,%
\begin{equation}
\lim_{t\rightarrow\infty}K\left(  t\right)  =\lim_{t\rightarrow\infty}%
\frac{2\sin\left(  ck_{0}-n\omega\right)  t}{ck_{0}-n\omega}=2\pi\delta\left(
ck_{0}-n\omega\right)  , \label{4.34}%
\end{equation}
see, e.g., \cite{SokolovTernov}. Taking Eq. (\ref{4.34}) into account and the
fact that the delta-function in the RHS of Eq. (\ref{4.34}) vanishes for
negative $n$, we obtain:
\begin{equation}
\lim_{t\rightarrow\infty}w\left(  t\right)  =\frac{q^{2}\omega^{2}}{c}%
\sum_{n=1}^{+\infty}n^{2}\int_{0}^{\pi}\sin\theta\left[  j_{n}^{2}\left(
\frac{n\omega R}{c}\sin\theta\right)  \cot^{2}\theta+\frac{\omega^{2}R^{2}%
}{c^{2}}j_{n}^{\prime2}\left(  \frac{n\omega R}{c}\sin\theta\right)  \right]
d\theta. \label{4.35}%
\end{equation}
The result (\ref{4.35}) reproduces literally the Schott formula for the rate
of the energy radiation by a classical current.

\subsection{Schwinger calculations of the one-photon radiation}

Schwinger in his work \cite{Schwi49} considered classical SR, using the method
is based on an examination of the energy transfer rate from the electron to
the electromagnetic field. Later in Ref.~\cite{Schwi54} he calculated the
quantum corrections of the first order in $\hbar$ to the classical formula,
taking into account the quantum nature of the radiating particle but
neglecting its spin properties. In 1973 he reexamined the problem, utilizing
the source theory to obtain the quantum expression for the spectral
distribution of the radiated power \cite{Schwi73}.

In \cite{Schwi49}, he presented several different distributions of the
instantaneous power. Among them is expression for the power radiated into a
unit solid angle about the direction $\mathbf{n}=\left(  \cos\varphi\cos
\theta,\sin\varphi\cos\theta,\sin\theta\right)  $ and contained in a unit
angular frequency interval about the frequency $ck_{0}$,
\begin{align}
&  P\left(  \mathbf{n},k_{0}\right)  =\sum_{n=1}^{\infty}\delta\left(
ck_{0}-n\omega\right)  P_{n}\left(  \mathbf{n}\right)  ,\nonumber\\
&  P_{n}\left(  \mathbf{n}\right)  =\frac{\omega^{2}R}{c^{2}}\frac{q^{2}}%
{2\pi}n^{2}\left[  \frac{\omega^{2}R^{2}}{c^{2}}j_{n}^{\prime2}\left(
\frac{n\omega R}{c}\cos\theta\right)  +\frac{\sin^{2}\theta}{\cos^{2}\theta
}j_{n}\left(  \frac{n\omega R}{c}\cos\theta\right)  \right]  . \label{6.1}%
\end{align}
The total radiated power can be calculated as%
\begin{equation}
P=\int_{0}^{\infty}cdk_{0}\int P\left(  \mathbf{n},k_{0}\right)  d\Omega.
\label{6.2}%
\end{equation}
Considering the high-frequency radiation,%
\begin{equation}
1-\frac{\omega^{2}R^{2}}{c^{2}}\ll1,\ \theta\ll1,\ n\gg1, \label{6.3}%
\end{equation}
and using the connection between the Airy and Bessel functions, Schwinger
obtained an alternative representation for his result in the form%
\begin{align}
&  P_{n}\left(  \mathbf{n}\right)  =\frac{q^{2}\omega}{6\pi^{2}R}n^{2}\left(
1-\omega^{2}R^{2}/c^{2}+\theta^{2}\right)  ^{2}\left[  K_{2/3}^{2}\left(
\zeta\right)  +\frac{\theta^{2}K_{1/3}^{2}\left(  \zeta\right)  }{1-\omega
^{2}R^{2}/c^{2}+\theta^{2}}\right]  ,\nonumber\\
&  \zeta=\frac{n}{n_{c}}\left(  \frac{1-\omega^{2}R^{2}/c^{2}+\theta^{2}%
}{1-\omega^{2}R^{2}/c^{2}}\right)  ^{3/2}, \label{6.4}%
\end{align}
and $n_{c}$ is a critical harmonic number \cite{Schwi49}. Note that formal
difference in angular distribution between (\ref{6.1}) and (\ref{4.35}) appear
due to different notation and does not lead to any differences in final values.

In Ref.~\cite{Schwi54} he considered the quantum corrections of the first
order in $\hbar$ to the classical formula, taking into account the quantum
nature of the radiating electron. In his consideration he neglected the spin
properties as at this level of accuracy, the spin degrees of freedom play no
role for unpolarized particles. The first-order in $\hbar$ correction to the
classical formula (\ref{6.2}) can be obtained from the classical expression
for the differential radiation probability $\left(  ck_{0}\right)
^{-1}P\left(  \mathbf{n},ck_{0}\right)  $ \cite{Schwi54} by making the
substitution
\begin{equation}
ck_{0}\rightarrow ck_{0}\left(  1+\frac{\hbar ck_{0}}{E}\right)  .
\label{6.12}%
\end{equation}
The total radiated power with the first order quantum corrections obtained by
Schwinger reads
\begin{equation}
w=\frac{2}{3}\omega\frac{q^{2}}{R}\left(  \frac{E}{mc^{2}}\right)  ^{4}\left[
1-\sqrt{3}\frac{55}{16}\frac{\hbar}{mcR}\left(  \frac{E}{mc^{2}}\right)
^{2}+O\left(  \hbar^{2}\right)  \right]  . \label{6.13}%
\end{equation}

In Ref. \cite{Schwi73} Schwinger considered the radiation of a spinless
charged particle in the homogeneous magnetic field, and obtained the spectral
distribution of the radiated power $w\left(  k_{0}\right)  $ (here $c=\hbar
=1$) in the form%
\begin{equation}
w\left(  k_{0}\right)  =\frac{ck_{0}q^{2}}{\pi m}\frac{m^{2}}{E^{2}}\left\{
\int_{0}^{\infty}\frac{dx}{x}\left(  1+2x^{2}\right)  \sin\left[  \frac
{ck_{0}}{\omega}\left(  \frac{m}{E}\right)  ^{3}\left(  x-\frac{x^{3}}%
{3}\right)  \right]  -\frac{1}{2}\pi\right\}  ,\ x=\frac{1}{2}\omega t\frac
{E}{m}. \label{6.14}%
\end{equation}
According to the author, Eq.\ (\ref{6.14}) in the classical limit reproduces
the Schott formula.

Note that the formulas (\ref{7.1}) and (\ref{6.14}) include both the
corrections due to electron recoil and the effects of quantization of the
electromagnetic field. As for the comparison with our result, the angular
distributions coincide with the Schott formula and are not affected by quantum corrections.

\subsection{One-photon radiation of scalar particles due to transitions
between Landau levels}

When presenting the results obtained by other authors, we use the same system
of units that was utilized in the cited articles.

There is a different approach to calculation of radiation of the spinless
charged particle in due to one-photon transitions between the energy levels
presented in Ref.~\cite{scalar}. These calculations are based on the exact
solutions of the Klein-Gordon equation in the uniform magnetic field (the
Furry picture approach). The spectral angular distribution of the radiated
power in this approach has the form%
\begin{align}
&  w=\frac{27}{16\pi^{2}}w_{0}\xi^{2}\varepsilon_{0}^{-5/2}\int_{0}^{\infty
}dy\int_{0}^{\pi}\frac{\sin\theta d\theta}{\left(  1+\xi y\right)  ^{3}}%
y^{2}\left[  \varepsilon^{2}K_{2/3}^{2}\left(  z_{0}\right)  +\varepsilon
\cos\theta K_{1/3}^{2}\left(  z_{0}\right)  \right]  ,\nonumber\\
&  w_{0}=\frac{8}{27}\frac{q^{2}m^{2}c^{2}}{\hbar^{2}},\ \xi=\frac{3}{2}%
\frac{e\hbar H}{m^{2}c^{3}}\frac{E}{mc^{2}},\ \varepsilon_{0}=\left(
\frac{mc^{2}}{E}\right)  ^{2},\nonumber\\
&  z_{0}=\frac{y}{2}\left(  \frac{\varepsilon}{\varepsilon_{0}}\right)
^{3/2},\ \varepsilon=1-\frac{\omega^{2}R^{2}}{c^{2}}\sin^{2}\theta
,\ E=\frac{mc^{2}}{\sqrt{1-\omega^{2}R^{2}/c^{2}}}, \label{7.1}%
\end{align}
where $K_{n}\left(  z_{0}\right)  $ are Airy functions, and $E$ is the
electron energy. Unfortunately, no representation of the (\ref{7.1}) in terms
of the Bessel functions is given by the authors; however, it is claimed that
Eq.~(\ref{7.1}) in the limit\textrm{ }$\hbar\rightarrow0$ reproduces the
classical result.

\section{Two-photon radiation}

The probability $p\left(  2,t\right)  $ and the energy $W\left(  2,t\right)  $
of the two-photon radiation for a circular current (\ref{4.1}) have the form:%
\begin{align}
&  p\left(  2,t\right)  =\frac{\alpha^{2}}{\left(  2\pi\right)  ^{2}}\left\{
\int\frac{d\mathbf{k}}{2k_{0}}\left[  k_{0}^{-2}F_{1}\left(  \mathbf{k}%
,t\right)  \cot^{2}\theta+R^{2}F_{2}\left(  \mathbf{k},t\right)  \right]
\right\}  ^{2},\nonumber\\
&  W\left(  2,t\right)  =\frac{\alpha^{2}\hbar c}{\left(  2\pi\right)  ^{2}%
}\left\{  \int d\mathbf{k}\left[  k_{0}^{-2}F_{1}\left(  \mathbf{k},t\right)
\cot^{2}\theta+R^{2}F_{2}\left(  \mathbf{k},t\right)  \right]  \right\}
\nonumber\\
&  \times\left\{  \int\frac{d\mathbf{k}^{\prime}}{k_{0}^{\prime}}\left[
k_{0}^{\prime-2}F_{1}\left(  \mathbf{k}^{\prime},t\right)  \cot^{2}%
\theta^{\prime}+R^{2}F_{2}\left(  \mathbf{k}^{\prime},t\right)  \right]
\right\}  , \label{9.1}%
\end{align}
where%
\begin{equation}
F_{1}(\mathbf{k},t)=\left\vert \sum_{n=-\infty}^{+\infty}nj_{n}\left(
k_{\perp}R\right)  F_{\mathbf{k}}^{n}\left(  \varphi,t\right)  \right\vert
^{2},\ F_{2}(\mathbf{k},t)=\left\vert \sum_{n=-\infty}^{+\infty}j_{n}^{\prime
}\left(  k_{\perp}R\right)  F_{\mathbf{k}}^{n}\left(  \varphi,t\right)
\right\vert ^{2}. \label{9.3}%
\end{equation}

It is useful to compare our results with the calculations of two-photon
radiation presented in other works. In the Ref. \cite{Zhuk76}, it was
considered the bremsstrahlung of relativistic electrons in the so-called
approximation of soft photons (the total energy of emitted photons is much
less than the energy of a relativistic electron). Our initial assumption, that
the classical current $J(x)$ remains unchanged, despite the radiation losses
matches with this approximation. The authors of Ref. \cite{Zhuk76} had used
the expression for the instantaneous spectral distribution of the radiation
energy of an electron using the Liénard-Wiechert potentials. In such a way
they have obtained the total electromagnetic energy of the one-photon
radiation. If the electric current in the latter quantity is taken in the form
(\ref{4.1}), it coincides with our result $W\left(  1,t\right)  $ given by Eq.
(\ref{4.24}). Then the probability of emitting a photon is defined by the
authors as $p\left(  \left\{  1\right\}  ,t\right)  =W\left(  \{1\},t\right)
/\left(  \hbar ck_{0}\right)  $ [here $W\left(  \{1\},t\right)  $ is the
integrand of $W\left(  1,t\right)  $] and the probability $p\left(  \left\{
N\right\}  ,t\right)  $ of emitting $\left\{  N\right\}  $ soft photons in a
narrow range of angles along the electron motion direction reads:
\begin{equation}
p\left(  \left\{  N\right\}  ,t\right)  =\prod_{i=1}^{N}p\left(
1_{\mathbf{k}_{i}\lambda_{i}},t\right)  =\prod_{i=1}^{N}\left\vert
y_{\mathbf{k}_{i}\lambda_{i}}\right\vert ^{2}. \label{9.5}%
\end{equation}
According to the authors, "when integrating in a finite interval of
frequencies and directions, one must to introduce a factor $\left(  N!\right)
^{-1}$ that takes into account the identity of the photons".\textrm{ }Thus,
they arrive to our result (\ref{8.2a}), which contains such factor for any
momenta $\mathbf{k}$ without heuristic prescriptions. It is easy to verify
that using the same approximation of the small difference between the angles
$\varphi_{1}$ and $\varphi_{2}$ of photons emitted, $\Delta\varphi=\left(
\varphi_{1}-\varphi_{2}\right)  \ll1$, we obtain from Eq. (\ref{8.2a}) for the
probability of the two-photon radiation the following result:%
\begin{equation}
p\left(  2,t\right)  =\frac{25}{24}\alpha^{2}\omega\gamma\Delta\varphi
,\ \gamma=\left(  1-\omega^{2}R^{2}/c^{2}\right)  ^{-1/2}. \label{9.6}%
\end{equation}
It coincides with the one of the work \cite{Zhuk76}.

It should be noted that in Refs. \cite{DualPhot} and \cite{MultiPhot}, the
authors calculated two-photon synchrotron emission, considering electron
transitions between Landau levels by the help of the corresponding solutions
of the Dirac equation. In the approximation accepted in the work \cite{Zhuk76}
they derived corrections to Eq. (\ref{9.6}) of the order $\hbar$ due to the
quantum nature of the electron and due to its spin.

\section{Concluding remarks\label{S5}}

As was said in the Introduction, in the beginning, the SR was studied by
classical methods using the Liénard-Wiechert potentials of electric currents.
Subsequently, it became clear that in some cases, quantum corrections to
classical results may be important. These corrections were studied,
considering the emission of photons arising from electronic transitions
between spectral levels, described in terms of the Dirac equation. In this
paper, we have considered an intermediate approach, in which electric currents
generating the radiation are treated classically, whereas the quantum nature
of the radiation is taken into account exactly. Such an approximate approach
allows one to study the one-photon and multi-photon radiation without
complicating calculations using corresponding solutions of the Dirac equation.
We have constructed exact quantum states (\ref{2.10}) of the electromagnetic
field interacting with classical currents and studied their properties. By
their help, we have calculated a probability of photon emission by classical
currents from the vacuum initial state and obtained relatively simple general
formulas for the one-photon and multi-photon radiation. Using the specific
circular electric current, we have calculate the corresponding one-photon and
two-photon SR. It was demonstrated that the emitted single-photon power per
unit time in the limit $t\rightarrow\infty$ coincides with the classical
expression obtained by Schott. This is not strange, since Schott's result was
already semi-classical, since he treated the electromagnetic field in terms of
the Maxwell's equations. It is well known that, see e.g. \cite{AkhiBer81}, in
fact, the Maxwell equations can be interpreted as the Schrödinger equation for
a single photon, the absence of the Planck constant $\hbar$ in these equations
as well as in the Schott formula is associated with the masslessness of the
photon. The consideration of the electromagnetic radiation in a semiclassical
manner, using Maxwell's equations, often allows one to study quantum effects
the of radiation \cite{comparison}. Schwinger's calculations of SR contain
$\hbar$ since he used elements of QFT that take into account quantum character
of electron motion and in the limit $\hbar\rightarrow0$ lead to the Schott
result. The same situation takes place with calculations of the SR radiation
of a spinless charged particle due to transitions between energy levels with
one-photon emission presented in Ref.~\cite{scalar}. The proposed approach
provides an opportunity to separate the effects of radiation associated with
the quantum nature of the electromagnetic field from the effects caused by the
quantum nature of the electron. The calculation of multiphoton corrections is
significantly simplified compared, for example, with the approach described in
\cite{MultiPhot, DualPhot,Zhuk76}, where a two-photon correction to the
radiation of an electron moving in a circular orbit in a constant uniform
magnetic field is calculated within the framework of the Furry picture.
Finally, it becomes possible to study the initial states of the system other
than the vacuum initial state (the state without initial photons). Using these
state vectors, the probabilities $p\left(  N,t\right)  $ (\ref{8.10}) and the
energy $W\left(  N,t\right)  $ (\ref{8.13}) of $N$ photon radiation induced by
classical currents are derived. The latter quantity can be summed exactly
representing the total energy $W\left(  t\right)  $ (\ref{8.15}) of emitted
photons. The obtained results can be used for the systematic study of the
multiphoton SR.

\section{Acknowledgements}

Bagrov acknowledges support from Tomsk State University Competitiveness
Improvement Program. Gitman is supported by the Grant No. 2016/03319-6,
Fundação de Amparo à Pesquisa do Estado de São Paulo (FAPESP), and permanently
by Conselho Nacional de Desenvolvimento Científico e Tecnológico (CNPq). The
work of Shishmarev was supported by the Russian Foundation for Basic Research
(RFBR), project number 19-32-60010.

\section*{Appendix}

\appendix\setcounter{equation}{0} \renewcommand{\theequation}{A\arabic{equation}}

Here we show that the sum (\ref{8.14}) can be calculated analytically with the
help of representation (\ref{8.12}). We start at the definition of $W\left(
N,t\right)  $ from Eq.~(\ref{8.12}),%
\begin{equation}
W\left(  N,t\right)  =\hbar c\left(  N!\right)  ^{-1}\sum_{\lambda_{1}=1}%
^{2}\sum_{\lambda_{2}=1}^{2}\ldots\sum_{\lambda_{N}=1}^{2}\int d\mathbf{k}%
_{1}d\mathbf{k}_{2}\ldots d\mathbf{k}_{N}\left[  \sum_{j=1}^{N}\left\vert
\mathbf{k}_{j}\right\vert \right]  \prod_{i=1}^{N}\left\vert y_{\mathbf{k}%
_{i}\lambda_{i}}\right\vert ^{2}. \label{ap3.1}%
\end{equation}
\textrm{ }We first consider the term with $j=1$. In the entire integrand
(\ref{ap3.1}), only the factor $\left\vert \mathbf{k}_{1}\right\vert
\left\vert y_{\mathbf{k}_{1}\lambda_{1}}\right\vert ^{2}$ depends on
$\lambda_{1}$ and\textrm{ }$\mathbf{k}_{1}$. Therefore, everything except the
factor $\left\vert \mathbf{k}_{1}\right\vert \left\vert y_{\mathbf{k}%
_{1}\lambda_{1}}\right\vert ^{2}$ can be taken out from the signs of the sum
over $\lambda_{1}$ and the integral over $d\mathbf{k}_{1}$. Since the indices
$i$ are dumb (the limits of all summations and integrations are the same), we
can cyclically shift their numbering ($i\rightarrow i-1$, i.e., $2\rightarrow
1$, $3\rightarrow2$, \ldots,\ $N\rightarrow N-1$,\ $1\rightarrow N$). We do
the same with each term from the sum $j=2,3,4,\ldots,N-1$. Now it's obvious
that the sum over $j$ in (\ref{ap3.1}) degenerates into a factor $N$, and the
quantity $W\left(  N,t\right)  $ takes the form:%
\begin{equation}
W\left(  N,t\right)  =\frac{\hbar c}{\left(  N-1\right)  !}\sum_{\lambda
_{1}=1}^{2}\sum_{\lambda_{2}=1}^{2}\ldots\sum_{\lambda_{N}=1}^{2}\int
d\mathbf{k}_{1}d\mathbf{k}_{2}\ldots d\mathbf{k}_{N}\left\vert \mathbf{k}%
_{N}\right\vert \prod_{i=1}^{N}\left\vert y_{\mathbf{k}_{i}\lambda_{i}%
}\right\vert ^{2}. \label{ap3.4}%
\end{equation}
It is easy to see that Eq. (\ref{ap3.4}) can be written as:
\begin{equation}
W\left(  N,t\right)  =\frac{\hbar c}{\left(  N-1\right)  !}\sum_{\lambda
_{N}=1}^{2}\int d\mathbf{k}_{N}\left\vert \mathbf{k}_{N}\right\vert \left\vert
y_{\mathbf{k}_{N}\lambda_{N}}\right\vert ^{2}\prod_{i=2}^{N}\left[
\sum_{\lambda_{i}=1}^{2}\int d\mathbf{k}_{i}\left\vert y_{\mathbf{k}%
_{i}\lambda_{i}}\right\vert ^{2}\right]  . \label{ap3.5}%
\end{equation}
Finally, getting rid of dumb indices, we obtain:%
\begin{align}
&  W\left(  N,t\right)  =\frac{\hbar cA}{\left(  N-1\right)  !}\left[
\sum_{\lambda=1}^{2}\int d\mathbf{k}\left\vert y_{\mathbf{k}\lambda
}\right\vert ^{2}\right]  ^{N-1},\nonumber\\
&  A=\sum_{\lambda=1}^{2}\int d\mathbf{k}k_{0}\left\vert y_{\mathbf{k}\lambda
}\right\vert ^{2},\ k_{0}=\left\vert \mathbf{k}\right\vert . \label{ap3.6}%
\end{align}
The total energy $W\left(  t\right)  $ reads:%
\begin{equation}
W\left(  t\right)  =\sum_{N=1}^{\infty}W\left(  N,t\right)  =\hbar
cA\sum_{N=1}^{\infty}\left[  \left(  N-1\right)  !\right]  ^{-1}\left[
\sum_{\lambda=1}^{2}\int d\mathbf{k}\left\vert y_{\mathbf{k}\lambda
}\right\vert ^{2}\right]  ^{N-1}. \label{ap3.7}%
\end{equation}
The sum over $N$ can be reduced to an exponent by the change $N=M-1$. Thus, we
justify Eq. (\ref{8.15}).

\end{document}